\documentclass[10pt,letterpaper,twocolumn]{article} % single column, double spaced
\usepackage{ol2}
\usepackage{epstopdf}
\usepackage{amsmath}

\newcommand{\einh}[1]{\,\mathrm{#1}}

\begin{document}

\twocolumn[ %% activate for two-column option

\title{Fabrication and characterization of an electrically contacted vapor cell}

%% For REVTeX it is possible to automate superscript and e-mail callouts with the superscriptaddress option; see REVTeX4 documentation.

\author{R. Daschner$^1$, R. Ritter$^1$, H. K\"ubler$^1$, N. Fr\"uhauf$^2$, E. Kurz$^2$, R. L\"ow$^1$, and T. Pfau$^{1,*}$}

\address{
$^1$5. Physikalisches Institut, Universit\"at Stuttgart, Pfaffenwaldring 57, 70550 Stuttgart, Germany \\
$^2$Institut f\"ur Gro\ss fl\"achige Mikroelektronik, Universit\"at Stuttgart, 70569 Stuttgart, Germany \\
$^*$Corresponding author: t.pfau@physik.uni-stuttgart.de
}

\begin{abstract}
We demonstrate the use of electrically contacted vapor cells to switch the transmission of a probe laser. The excitation scheme makes use of electromagnetically induced transparency involving a Rydberg state. The cell fabrication technique involves thinfilm based electric feedthroughs which are well suited for scaling this concept to many addressable pixels like in flat panel displays.

\end{abstract}

%\ocis{310.7005, 020.5780, 020.6580}

 ] %% activate for two-column option

\noindent
Saturated atomic vapors at room temperature are the basis for applications such as atom clocks \cite{kna2005}, magnetic field sensors \cite {polzik2010} or optical frequency references \cite{DAVLL}. New applications in nonlinear and quantum optics are enabled by the possibility to coherently excite Rydberg states in thermal vapors \cite{Adams2007,loe2009}. As Rydberg states show a huge polarizability and a correspondingly large Stark effect the optical properties of media involving such states can be switched and modulated by moderate electric fields \cite{mohapatra2008}. In fact uncontrolled electric fields can lead to drifts, broadenings and dephasing. While electric field control is not a fundamental problem in vacuum cells with macroscopic feedthroughs, conventional glass vapor cells are sealed above the melting point of glass which does not allow for thinfilm based feedthroughs. Such electric feedthroughs can be scaled up to larger numbers for example to address many pixels simultaneously like in flat panel displays. Therefore to convert Rydberg based optical media into actual scalable and miniaturized devices, it is necessary to develop a technique  to produce vacuum tight vapor cells with thinfilm based feedthroughs which can be produced in large scale with well established methods used in industry. Miniaturized cells compatible with silicon platforms \cite{schmidt2010} and microscopic glass based vapor cells \cite{balu2010} have been fabricated without electric field control. Miniaturization of a vapor based atom clock  ended up as a commercial product \cite{kna2005}.

%In addition such cells should be easily integrable with the typical platforms and also try to reduce power consumption as much as possible. To satisfy these requirements it is typically advantageous to miniaturize the key elements. 

Here we adopt the highly developed and standardized technologies from liquid crystal display (LCD) fabrication to produce vacuum tight cells to be filled with Rubidium vapor including electric field control. Large scale integrated field plates, pixels and feedtroughs to address liquid crystal media, is a standard technology in LCDs. We show that a similar approach can be used to fabricate vacuum tight vapor cells for reactive alkali atoms like Rubidium. The materials to be used have to fulfill four fundamental requirements: 1) compatibility with a pressure below $10^{-4}$ mbar 2) high transparency for the excitation light 3) conductive and 4) inert against chemical reactions with Rubidium atoms. In our approach we use borosilicate glass as a basic material, which fulfills almost perfectly the requirements 1, 2 and 4. To add also transparent and at the same time conductive layers we chose two realizations. The first one uses thin enough metallic layers below $10\einh{nm}$ of Ni and Al, which still maintain a large enough transparency of 75\% and 65\%, respectively, without acquiring a too large resistance. The second uses non-inert ITO (IndiumTinOxide) layers, which are protected against reactions with Rubidium by a thin dielectric coating.

%\textcolor{red}{Roberts Baustellen: Unterscheidung classical und quantum devices, protective coatings a la polzik, Billig, large scale processes, bekannte %techniken, klein, wenig power consumption. Messung zeigt Vakuum besser als \"uber eine Zeit von 20 to 40 days depending on coating , Scaleability des %Ansatzes im Outlook, Netwzerk via Photonen, oder WW durch glasw\"ande hindurch.}

\begin{figure}
a)\\
\centerline{\includegraphics[width=8.4cm]{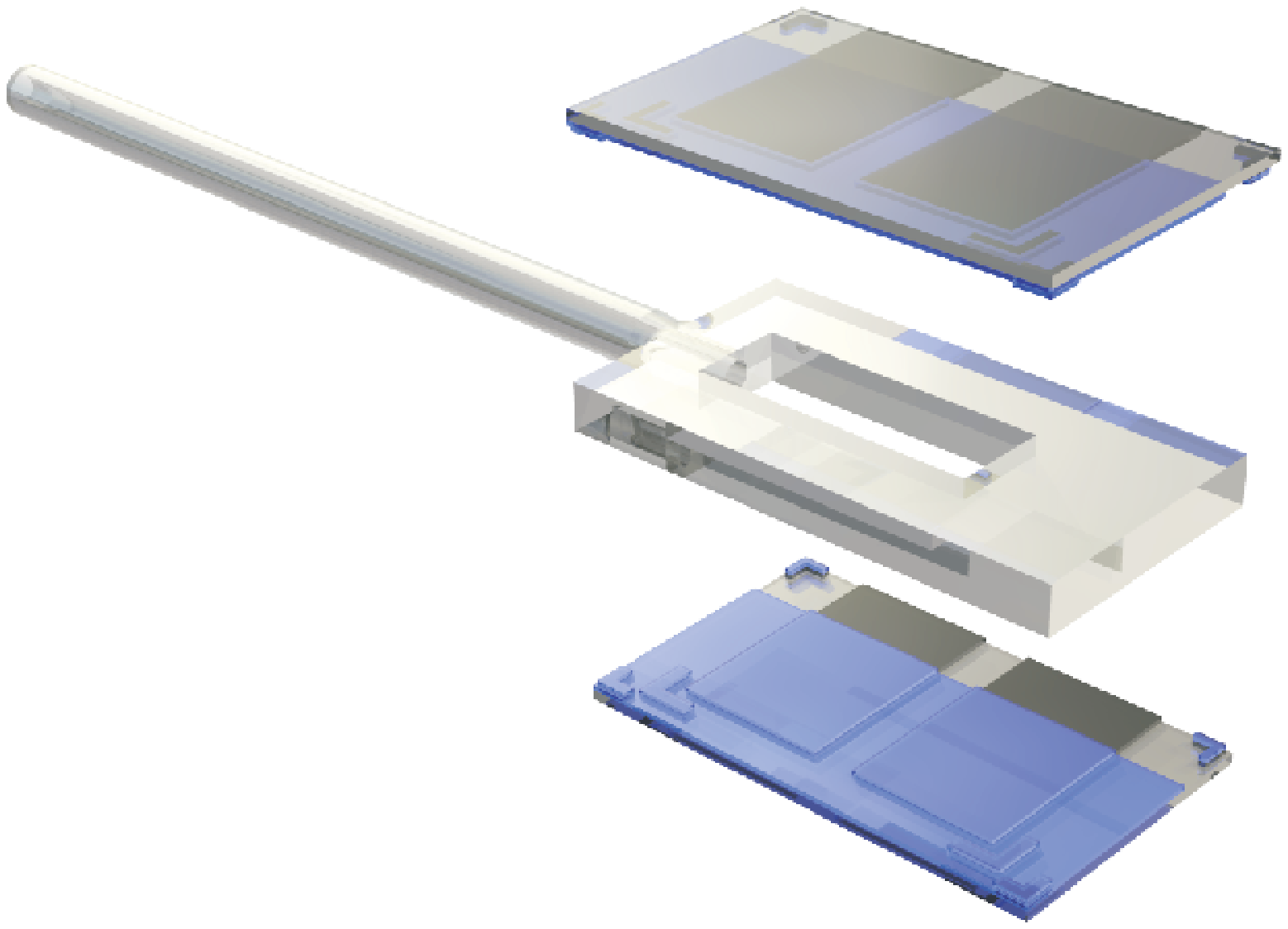}}\\
b)\\
\centerline{\includegraphics[width=8.4cm]{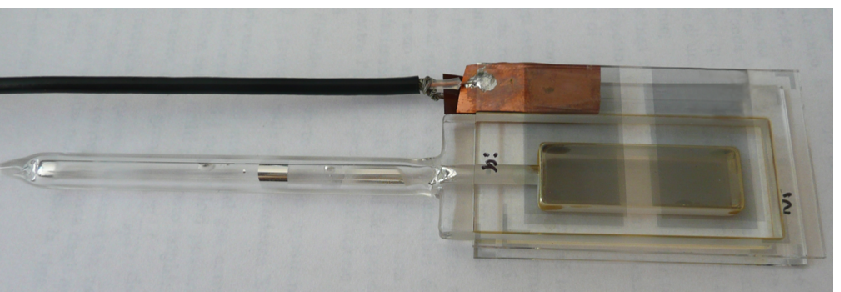}}
\caption{(Color online) a) Scheme: A glass frame provides a volume which is closed by glueing two slides with thinfilm conductive elements. The Rb reservoir is attached to the frame. b) Electrically contacted vapor cell with Ni-electrodes. The electrodes have been structured into four field plates with an area of $10\einh{mm}\,\times\,12.5\einh{mm}$ each to represent two independent pixels. Each of the electrodes is contacted to the outside. The tube on the left side is filled with Rubidium and serves as a reservoir. 
\label{cell_picture}}
\end{figure}

Our blank cells are assembled in two steps. First we melt a frame with a thickness of $5\einh{mm}$ radially to a small tubing, which will later serve as a reservoir for the atoms. In the second step we use a vacuum tight epoxy to attach on both sides of the frame a $1.1\einh{mm}$ thick glass plate (see Fig. \ref{cell_picture} a)) since almost any additional thin layer coating on the glass plates would not survive the welding temperatures for glass.
The size of the coverslips extends over the dimensions of the frame which allows for a connection of the conductive structures to the outside as shown in Fig \ref{cell_picture} b). The different coatings are produced by sputtering the desired materials on the substrates which can be structured in a second step with photolithographic methods. In case of ITO ($100\einh{nm}$ thick) a dielectric protection layer of  $100\einh{nm}$  SiN or $200\einh{nm}$ of $\mathrm{SiO_x}$ is put on top via PECVD (plasma enhanced chemical vapor deposition) to avoid chemical reaction of Rubidium with ITO. To finally fill the cells with Rubidium we heat the coated and glued cells to $75\einh{^\circ C}$ and pump it, transfer a droplet of Rubidium into the cell and continue pumping for three days at $75\einh{^\circ C}$ to extract all possible gaseous reaction products of the Rubidium and the epoxy.  After sealing the reservoir we can adjust the vapor pressure inside the cell by heating the reservoir and the cell to moderate temperatures up to $120\einh{^\circ C}$. 

%Due to the higher conductivity of metals the layer thickness for metal coated cells can be reduced to below $10\einh{nm}$ resulting in a resistivity of $285\einh{\Omega}$ and $117\einh{\Omega}$ per square for Ni and Al respectively. In addition to that no protection layer is needed but the transmission is only 75\% for Ni and 65\% for Al. 

The performance of the cells and the quality of the applied electric field is probed by electromagnetically induced transparency (EIT) spectroscopy involving Rydberg states in a three level ladder system \cite{Adams2007}. The first excitation step at $795\einh{nm}$ is resonant to the $5\mathrm{S}_{1/2}F=3\rightarrow5\mathrm{P}_{1/2}F=2$ transition in $\mathrm{^{85}Rb}$ and acts as the probe field. The second laser at $474\einh{nm}$ represents the pump field and is scanned with respect to the 5P$_{1/2}$ to 32S transition. In the end we infer from the measured widths of the EIT-peaks the coherence times. %and from the positions of the resonances the actual electric fields inside the cell.

Different AC and DC voltages are applied to one of the pixels and the Stark shift of the Rydberg level is observed by determining the position of the EIT relative to the one in a reference spectroscopy cell from which one can determine the actual electric fields inside the cell.

%The pixels can be addressed individually and thereby an electric field can be applied along the direction the lasers pass through the cell. 
The cell and the reservoir are heated by two independent ovens as described in \cite{kue2010}, allowing a good control over the vapor pressure inside the cell.

\begin{figure}
\centerline{\includegraphics[width=8.4cm]{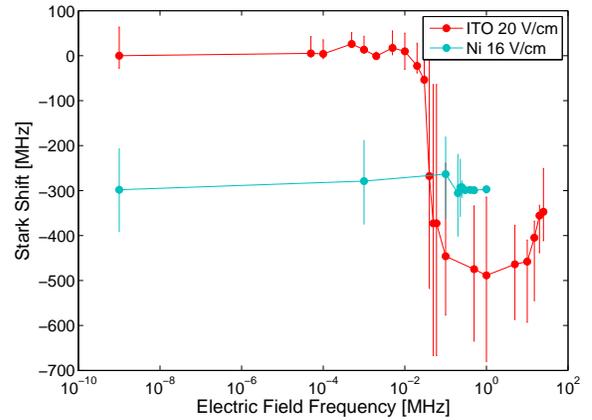}}
\caption{(Color online) Frequency dependence of the Stark shift in the ITO and the Ni coated cell. Due to the dielectric protection coating of the ITO ions inside the ITO cell shield the applied electric field for frequencies below 40 kHz. In the Ni coated cell the inner surface is conductive thus the ions can be drained away. \label{frequenzabh}}
\end{figure}
Figure \ref{frequenzabh} shows the frequency dependence of the measured Stark shift for a constant amplitude of the electric field in the ITO and the Ni cell at $20$ and $16\einh{V/cm}$ respectively. In the Ni coated cell no frequency dependence is observed. For the ITO cell however, we get maximum shift at a frequency of $1\einh{MHz}$. Below a threshold of $40\einh{kHz}$ the EIT peak is almost unshifted due to the low conductivity of the dielectric surface. The protection layer is too thick to allow enough current from the ITO contact through the protection layer to the inner surface of the cell. If ions and electrons are produced inside the cell for example by collisions, these charges stick to the inner surfaces and cannot be drained away. Therefore they shield the applied electric field and thus the Rydberg levels experience no shift \cite{Adams2007}. 
The upper frequency limit results from the electronics driving the AC voltages and is therefore not limiting the switching time in principle.

\begin{figure}
\centerline{\includegraphics[width=8.4cm]{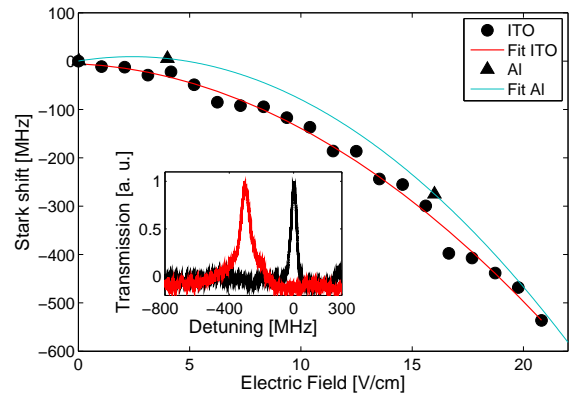}}
\caption{(Color online) The measured AC Stark shift in a cell coated with ITO/$\mathrm{SiO_x}$ and in a cell coated with Aluminium is almost the same. Inset: EIT signal at $0\einh{V_{pp}/cm}$ (black) and at $16\einh{V_{pp}/cm}$ (red) at a frequency of $500 \einh{kHz}$ in the Ni coated cell. The signal can be shifted by more than one linewidth, which is necessary to switch cells between ON (transmission) and OFF (no transmission). \label{AC_Stark_shift}
}
\end{figure}

In figure \ref{AC_Stark_shift} the position of the EIT peak is plotted over the applied AC electric fields between $0\einh{V_{pp}/cm}$ and $20\einh{V_{pp}/cm}$ with a modulation frequency of $1\einh{MHz}$. In the ITO as well as in the Al coated cell, we observe a quadratic shift. For the ITO cell the measured shift of $2.24\einh{MHz/(V/cm)^2}$ is close to the theory value of $2.26\einh{MHz/(V/cm)^2}$. To switch between transmission and absorption in a cell the EIT peak has to be shifted  by more than one linewidth. With a maximum linewidth of $90\einh{MHz}$ for the shifted EIT line this is easily achieved by applying $16\einh{V_{pp}/cm}$ (see inset of figure \ref{AC_Stark_shift}).

%\begin{figure}
%\centerline{\includegraphics[width=8.4cm]{shift_31_08}}
%\caption{(Color online) EIT signal at $0\einh{V_{pp}/cm}$ (black) and at $16\einh{V_{pp}/cm}$ (red) at a frequency of $500 \einh{kHz}$ in the Ni coated cell. The signal can be shifted by more than one linewidth, which is necessary to switch cells between ON (transmission) and OFF (no transmission). \label{switch}}
%\end{figure}

An array of cells should provide the possibility to switch the pixels independently. It is therefore necessary to apply DC electric fields to avoid crosstalk. As the protection layer of the ITO cell cannot be reduced enough to provide higher conductivities, this type of coating is not suitable for such an application.

\begin{figure}
\centerline{\includegraphics[width=8.4cm]{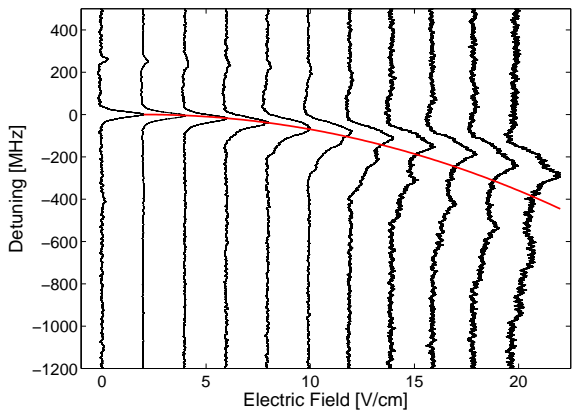}}
\caption{(Color online) EIT signals (black) and fitted Stark shift (red) in the Ni coated cell. For voltages above $12\einh{V/cm}$ we observe a broad distribution and a splitting of the EIT line due to ions produced in the cell. \label{DC_Stark_shift}}
\end{figure}
Metal coatings however (e.g. Al, Ni) allow for DC electric fields. EIT signals for different DC electric fields (see figure \ref{DC_Stark_shift}) also show a quadratic Stark shift. However at electric fields above $12\einh{V/cm}$ the line starts to broaden and split up to $200\einh{MHz}$ at $20\einh{V/cm}$. The 32S state we used in this experiment has no substructure that would split in electric fields. Also the first avoided crossing with high l-states for this state is at $25\einh{V/cm}$. The change of the lineshape could arise from different local electric fields originating from space charges inside the cell due to ionization of Rydberg atoms by collisions. These space charges build up an additional inhomogenous position dependent electric field which is overlapped with the applied field to an effective electric field. The summation over all EIT signals at different positions along the laser direction yields the measured signal.

To conclude we described a fabrication method for vapor cells, which provides the possibility to apply electric AC and DC fields inside the cell for a scalable number of pixels. Different materials for the thin film contacts have been studied, where for example ITO had to be protected by an additional dielectric layer to prevent chemical reactions with the Rubidium. Because of the accumulation of charges on that surface, only for frequencies above $40\einh{kHz}$ an effect could be observed. For the metal coating, where no protection layer is needed, also DC fields show a Stark effect on the Rydberg atoms. To probe this effect EIT measurements were performed and we showed, that we could shift the EIT resonance by more than a linewidth, allowing to change the pixel from an absorptive to a transparent state with a small applied electric field.

As a next step we plan to reduce the size and go to a higher number for the pixels. To do so, also the thickness of the cell will be reduced. Finally we envision to shrink the structures to a size of about 10$\einh{\mu m}$, which is the current technological standard for LCDs and in the order of the blockade radius for our Rydberg atoms.

We acknowledge the technical assistance of R. August and J. Quack. The work is supported by the ERC under contract number 267100, BMBF within QuOReP (Project 01BQ1013), the EU project MALICIA, DARPA Quasar program through a grant through ARO and contract research "Internationale Spitzenforschung II" of the Baden-W\"urttemberg Stiftung.

%\bibliographystyle{osajnl}
%\bibliography{verzeichnis1}{}

\pagebreak

\end{document}